# Estimation of the actual disease occurrence based on official case numbers during a COVID outbreak in Germany 2020

Ralph Brinks[1,2,3], Annika Hoyer[1]

(1) Department of Statistics, Ludwig-Maximilians-University Munich, Ludwigstr. 33, 80539 Munich, Germany
(2) Policlinics, Department and Hiller Research Unit for Rheumatology, Unversity Hospital Duesseldorf, Moorenstr. 5, 40225 Duesseldorf, Germany
(3) Institute for Biometry and Epidemiology, German Diabetes Center, Auf'm Hennekamp 65, 40225 Duesseldorf, Germany

## *Summary*

Since the beginning of March 2020, the cumulative numbers of cases of infection with the novel coronavirus SARS-CoV-2 in Germany have been reported on a daily basis. The reports originate from national laws, according to which positive test findings must be submitted to the Federal Health Authorities, the Robert Koch Institute, via the local health authorities. Since an enormous number of unreported cases can be expected, the question of how widespread the disease has been in the population cannot be answered based on these administrative reports. Using mathematical modeling, however, estimates can be made. These estimates indicate that the small numbers of diagnostic tests carried out at the beginning of the outbreak overlooked considerable parts of the infection. In order to cover the initial phase of future waves of the disease, wide-spread and comprehensive tests are recommended.

## *Introduction*

During the COVID outbreak in Germany in spring 2020, the cumulative number of cases of infected people was reported daily starting on March 4. The numbers of cases were based on the positive results reported according to the Infection Protection Act ("Infektionsschutzgesetz", IfSG) of the diagnostic tests carried out throughout Germany for the novel coranavirus SARS-CoV-2. The number of tests carried out was based on regional characteristics, availability of test kits and laboratory capacities. The frequency of the tests carried out increased during the epidemic [RKI 2020]. Together with the fact that several people with the associated COVID disease show only asymptomatic or mild courses and presumably have not been tested at any time, the variable frequency of the tests leads to the hypothesis that the actual disease process was at least partially underestimated.

Based on the administrative case numbers, in this work we provide a lower bound for the time course of the numbers of actually infected people. For this, we use the effective reproduction number ($R_{eff}$). In contrast to the actual number of cases, an estimate of $R_{eff}$ is possible if the cases are only reported incompletely [an der Heiden 2020]. Our resulting estimate of the lower bound of the numbers of infected people allows to further estimate an upper bound for the proportion of cases that were recorded in the administrative reports according to the IfSG. This proportion is referred to as the *case detection ratio* (CDR) [Borgdorf 2004].

## Methods

We start from the frequently used epidemiological model that Kermack and McKendrick described in their seminal work from 1927 [Kermack 1927]. Since the model consists of the three disease-relevant states 'Susceptible', 'Infectious' and 'Removed', i.e., recovered or dead, it is often called the SIR model [Vynnycky 2010]. The following ordinary differential equation applies to the number $I$ of infectious persons at time $t$

$$\frac{dI}{dt} = r \cdot I \cdot (R_{eff} - 1), \tag{1}$$

where $R_{eff}$ denotes the effective reproduction number. The effective reproduction number is the average number of secondary infectious cases that one primary case infects. The quantity $r$ in Equation (1) describes the rate at which people leave the stage of being infectious. A mathematical rationale for Equation (1) is given in the supplement to this work. Given the initial condition $I(0) = I_0$ at the starting time $t = 0$ of the official reports on March 4, the number of infectious persons at time $t$ can be derived from Equation (1). It holds:

$$I(t) = I_0 \cdot \exp\left( r \int_0^t \{R_{eff}(\tau) - 1\} d\tau \right). \tag{2}$$

As shown in the appendix to this work, the SIR model for the number $F_t$ of the newly occurring infectious cases (incident cases) in the period from $t$ to $t + \Delta$ leads to:

$$F_t = S(t) - S(t + \Delta) \approx R_{eff}(t) \cdot r \cdot I(t) \cdot \Delta. \tag{3}$$

Since the administrative case numbers were reported on a daily basis, we henceforth assume $\Delta = 1$ day.

Equations (1) to (3) apply to all infectious diseases that can appropriately be described using the SIR disease model [Vynnycky 2004, Chowell 2009]. These equations are independent of screening tests or any diagnostic procedures. By carrying out tests, the presence of the pathogens that underlie the disease becomes visible.

In the case of the novel coronavirus SARS-CoV-2, new infectious cases are reported daily via administrative channels. If in a reported case the date of onset of the disease has not been communicated, these new cases are assigned to a specific day using a statistical method [an der Heiden 2020]. Slight smoothing using a spline is used to smooth the number of reported incident cases.

Let the number of newly incident cases assigned to a day $t$ be denoted by $F_t^{(b)}$, we can consider the proportion of the reported cases in relation to the actual (incidental) new cases $F_t$ according to Equation (3). This leads to the case-detection ratio (CDR) [Borgdorf 2004]:

$$CDR_t = \frac{F_t^{(b)}}{F_t}. \tag{4}$$

In the case of a complete detection of all newly occurring infectious cases on one day $t$, the $CDR_t$ would equal 100%. Since in the early days of the outbreak of the SARS-CoV-2 epidemic only a few tests were carried out in Germany, it can be assumed that the *CDR* was low in early March.

Based on the reported reproduction number $R_{\text{eff}}$ and $r = 0.1$ per day [an der Heiden 2020], we use Equations (2) and (3) to calculate a lower bound for the number $I_0$ of infectious cases at time $t = 0$ we use that the *CDR* is at most 100% at any time. This means that the observed new cases $F_t^{(b)}$ cannot exceed the number of new cases $F_t$. For all times $t$ we have:

$$F_t^{(b)} = CDR_t \cdot F_t \leq F_t = R_{\text{eff}}(t) \cdot r \cdot \Delta \cdot I_0 \cdot \exp\left( r \int_0^t \{R_{\text{eff}}(\tau) - 1\} d\tau \right). \tag{5}$$

The only unknown quantity on the right side of (5) is the number $I_0$ of the persons infected at the time $t = 0$. As soon as we have a lower bound for $I_0$ and thus a lower bound for the number $F_t$ of the actual new cases via (5), we can determine an upper bound for the *CDR* using equation (4). As an illustration, we compare our estimate for the time course of the *CDR* with the number of positive tests performed.

## *Results*

Figure 1 shows the corrected new cases $F_t^{(b)}$ in the administrative reports of the Robert Koch Institute for the reporting period from March 4 to April 8 [an der Heiden 2020]. In addition, the smoothed curve is drawn as a black line.

The blue dashed curve in Figure 2 is a lower bound of the new cases $F_t$ based on the estimate (5). A lower bound of 7990 people is obtained for $I_0$.
If we now calculate the ratio $F_t^{(b)} / F_t$ for the determination of the upper bound of the CDR, we obtain the curve as depicted in Figure 3. This shows in particular that in the beginning of the epidemic in Germany the *CDR* was quite low with at most 30%.

In Figure 4, the quotient $F_t^{(b)} / F_t$ is compared to the number of positive results (number of tests performed multiplied by the positive rate) as red curve [RKI 2020]. Qualitatively, there is a similar course over time, which we see as an indication of our claim that the test behavior has changed over the reporting period.

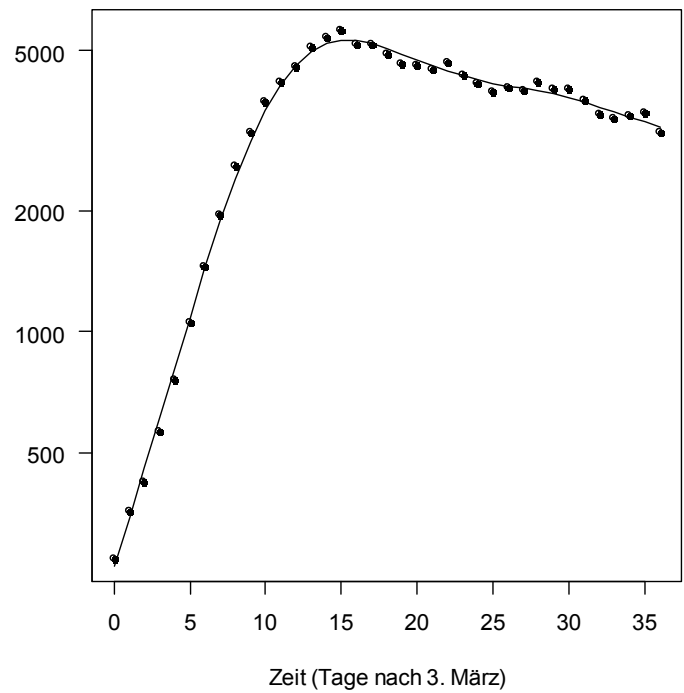

**Figure 1: Administrative numbers $F_t^{(b)}$ of reported incident cases of the SARS-CoV-2 infections in Germany (dots). The smoothed values are shown as solid line.**

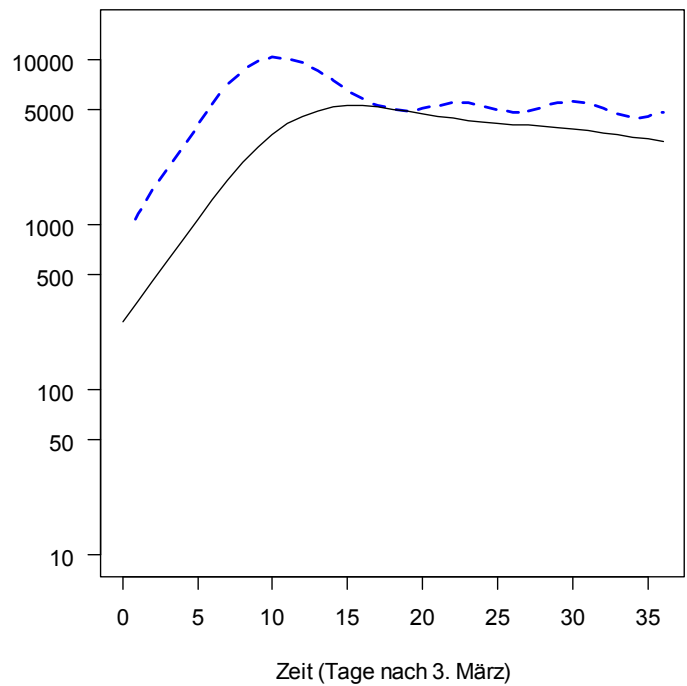

**Figure 2: Lower bound of the actual (true) numbers of incident cases $F_t$ (blue dashed line) in comparison with the reported numbers $F_t^{(b)}$ (black solid line).**

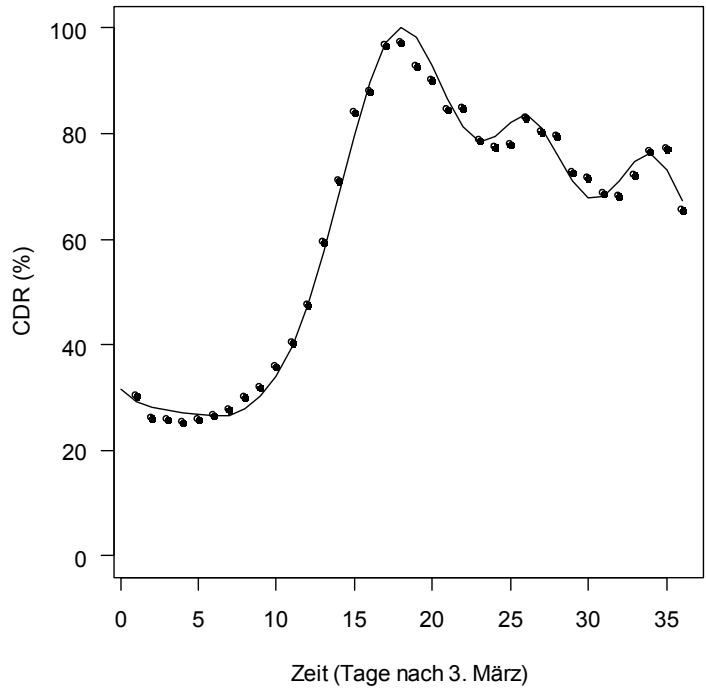

**Figure 1: Estimated upper bound of the case-detection-ratio (CDR in %) in Germany.**

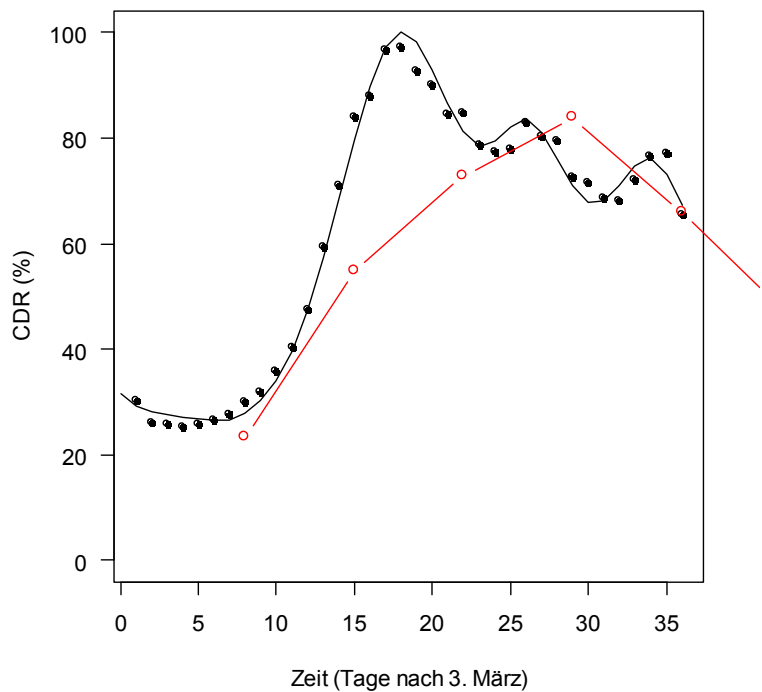

**Figure 4: Estimated upper bound of the CDR compared to the number of positive tests (red line).**

## Discussion

It is important to run screening and diagnostic tests to track the course of an epidemic outbreak and to assess possible measures to control the epidemic. In the extreme case that no tests are carried out, the epidemic goes widely unnoticed.

In this work we use the effective reproductive number $R_{eff}$ estimated over the generation time to estimate the actual occurrence of infections with the pathogen SARS-CoV-2 in the German population as a whole. A comparison of the reported cases ($F_t^{(b)}$) with the expected cases ($F_t$) gives an upper limit of the case detection ratio (*CDR*). As was shown on the basis of the daily reported case numbers, the initial phase of the SARS-CoV-2 outbreak in Germany remained largely undetected until around March 15.

At the start of reporting on March 4, a cumulative number of 262 cases was reported. Our estimate here shows that there were already around 8,000 infected people in Germany at this point in time. The phase of inadequate testing also includes the peak of infection around March 9, at which $R_{eff}$ reached the tentative maximum of 3.3. In order to reliably enable future surveillance of SARS-CoV-2, we need a close-knit recording with sufficient test capacities.

## Supplementary information

For giving the theoretical justification for Equations (1) and (3), we start with the SIR model [Chowell 2010, Vynnycky 2010], where *S*, *I*, *R* denote the numbers of susceptible, infectious and recovered people in the population, respectively. If the mortality of the susceptible people is negligible (for COVID it is), we have following system of differential equations

$$\frac{dS}{dt} = -\beta IS \qquad (A1)$$

$$\frac{dI}{dt} = \beta IS - rI \tag{A2}$$

$$\frac{dR}{dt} = rI. \tag{A3}$$

Here, $\beta$ is the rate of making an efficient contact (per time unit) and $r$ is rate at which infectious people leave the infectious state (due to recovery or death). Let $N = S + I + R$, then we have $\beta = R_0\, r\, /\, N$, where $R_0$ is the basic reproduction number [Vynnycky 2010]. Insertion into (A2) yields $dI/dt = r\, I\, (R_0\, S/N - 1)$. As $R_{\text{eff}} = R_0\, S/N$, we have Equation (1).

As $F_t = S(t) - S(t + \Delta)$ can be approximated by $-\Delta\, dS/dt$ (Taylor's Theorem), Equation (A1) and $\beta = R_0\, r\, /\, N$ yields $F_t \approx \Delta\, \beta\, I\, S = \Delta\, R_{\text{eff}}\, r\, I$, which is Equation (3).